\documentclass[12pt]{article} 
\usepackage[dvips]{graphics}
\title{Dual Phase Cosmic Rays}  
\author{{\it Richard Shurtleff~}\thanks{affiliation and mailing 
address: Department of Applied Mathematics and Sciences, 
Wentworth Institute of Technology, 550 Huntington Avenue, 
Boston, MA, USA, ZIP 02115, telephone number: (617) 989-4338, e-mail address: shurtleffr@wit.edu}} 
\begin{document} 
          
\maketitle 

\begin{abstract} A calculation based on flat spacetime symmetries shows how there can be two quantum phases. For one, extreme phase change determines a conventional classical trajectory and four-momentum, i.e. mass times four-velocity. The other phase occurs in an effective particle state, with the effective energy and momentum being the rate of change of the phase with respect to time and distance. A cosmic ray proton moves along a classical trajectory, but exists in an effective particle state with an effective energy that depends on the local gravitational potential. Assumptions are made so that a cosmic ray proton in an ultra-high energy state detected near the Earth was in a much less energetic state in interstellar space. A 300 EeV proton incident on the Earth was a 2 PeV proton in interstellar space. The model predicts such protons are in states with even more energy near the Sun than when near the Earth. 

PACS - 11.30.Cp, 11.10.-z, 04.60.-m, 96.50.S

Keywords - Poincare invariance, field theory, quantum gravity, cosmic rays

\end{abstract}

\section{Introduction}

In quantum mechanics, a particle's classical trajectory maximizes or minimizes the phase change along the trajectory. And the way the particle state's phase changes with location determines the particle's momentum and energy; e.g. the momentum is proportional to the change in phase per unit distance.

In Ref.\cite{S1}, hereafter referred to as `I', the phase that determines the trajectory and the phase that determines energy and momentum were found to differ in a way depending on the strength of the gravitational field. Thus one of these phases sends the particle along a classical trajectory while the other determines the particle state's energy and momentum. It is shown in this article that, for sufficiently high energy particles, even a weak gravitational potential can produce dramatic changes to the particle state compared to the state in a null potential.

In I a well-known method \cite{W} of obtaining the quantum fields of free massive particles is generalized. The field is constructed as a sum over particle creation and annihilation operators. The method relies on the group theory of spacetime symmetries and the generalizations involve translations. For free particles the momentum is unchanged by translation; that unfaithful representation (rep) assigns unity to all translations. By generalizing to a nontrivial rep, momentum can change with translation, so the motion in flat spacetime is in general curved. 

Having two distinct phases, `dual phases,' derives from a well known property of translations: any translation preserves all coordinate differences since any displacement cancels upon subtraction. The same coordinate differences occur when spacetime is translated through one displacement $\delta x$ and the particle state is translated through a second displacement $\epsilon.$

The process of constructing the quantum field constrains the displacement $\epsilon$ of particle states and their creation and annihilation operators. Suppose spacetime undergoes a Poincar\'{e} transformation. Any such transformation is equivalent to a boost-rotation combination $\Lambda$ followed by a displacement through $\delta x,$ which gives the coordinate transformation $x \rightarrow$ $\Lambda x + \delta x.$ 

When spacetime is transformed by $(\Lambda,\delta x),$ particle states and the operators transform by $(\Lambda,\epsilon),$ where the displacement $\epsilon$ is now a free parameter. In the construction of the quantum field, an expression is found for $\epsilon$ that involves a second rank tensor $M,$ which is shown in I to be related to the gravitational metric tensor $g_{\mu \nu},$ where $\mu,\nu \in$ $\{1,2,3,4\}$ = $\{x,y,z,t\}$ indicate Minkowski coordinates. One must know $g_{\mu \nu},$ $\Lambda,$ $x,$ and $\delta x$ to determine the displacement $\epsilon$ of the particle state. 

Perhaps unexpectedly, the particle state displacement $\epsilon$ is not homogeneous in the spacetime displacement $\delta x;$ there are terms proportional to $x.$ The particle state is translated even when spacetime is not and the transformation applied to the particle states depends on location whereas the transformation applied to spacetime and the quantum field does not depend on location. 

This puzzle is resolved as follows. Let $y$ denote the spacetime parameters of the particle states. Since the coefficient of $y$ in the phase is the momentum, the terms in $\epsilon$ proportional to $y$ contribute to an effective momentum $\bar{p}$ which replaces the eigenmomentum $p$ in the phase, $p \cdot y \rightarrow$ $\bar{p} \cdot y$ = $(p +\partial \epsilon/\partial y) \cdot y.$ Thus the different transformations applied to the particle states of a given momentum $p$ at different locations produce a single, common plane wave with an effective momentum $\bar{p}.$ The effective (four-)momentum therefore depends on $\Lambda,$ and on $g_{\mu \nu}$ through $M.$ We contend that the effective energy of the single, common plane wave is the energy observed by instruments that measure the energy dumped into the atmosphere by a cosmic ray.

Since $\epsilon$ depends on the Lorentz transformation $\Lambda$ and the trajectory does not, $\Lambda$ is free to vary. To see this, picture the trajectory as a sequence of translations in a given reference frame, i.e. $\delta x$s not $\Lambda$s. We choose the given frame to be determined by the distribution of Cosmic Microwave Background radiation. But to get to that given frame, one could start with any initial frame and transform to the given frame with a suitable $\Lambda.$ Since the initial frame is arbitrary, and since $\epsilon$ depends on $\Lambda,$ one must select an initial frame to determine the effective momentum.

We choose the initial frame to be the rest frame of the particle, see Fig. 1. The choice is based largely on its effect on the cosmic ray spectrum, but it should be noted that choosing the rest frame to be special is grounded in basic physical intuition.

With the rest frame as the initial frame, $\Lambda$ is a Lorentz transformation $L(p)$ taking the rest momentum $k$ = $\{0,0,0,m\}$ to the trajectory momentum $p.$ Then effective momentum $\bar{p}$ depends on the trajectory momentum $p$ and the gravitational metric tensor $g_{\mu\nu}.$ The trajectory momentum $p$ could be obtained by time-of flight methods and we assume that general relativity gives the gravitational metric tensor. It is shown in I that the trajectory momentum $p$ is consistent with the gravitational metric tensor $g_{\mu\nu}.$

Additional flexibility occurs by allowing the spacetime parameters $x$ for the quantum field and the trajectory to differ from the spacetime parameters $y$ of the particle states.  The construction of a quantum field involves the creation and annihilation operators, not the particle states directly. This leaves the spacetime of the particle states $y$ free to differ from $x$ by any Lorentz transformation $\lambda.$ In order to avoid having the spatial momentum $\overrightarrow{\bar{p}}$ decrease when the energy increases it is assumed that the spacetime of particle states $y$ is the time inversion of the spacetime $x$ of the quantum field  and the trajectory; $\lambda$ is the time inversion transformation.

With these assumptions, one finds that, for a null gravitational potential, $\phi$ = 0, the effective momentum $\bar{p}$ is just the  trajectory momentum $\bar{p}$ = $p.$ We are taking `momentum' to mean the four-momentum, so one component is the total energy, $\bar{p}^t$ = $m c^{2} + \bar{E},$ where $c$ is the speed of light and $m$ is the particle mass. 

For the purpose of illustration here, consider only high energy particles, $-\phi \gamma^2 >> $ 1, where gamma $\gamma m c^{2}$ = $m c^{2}+E$ and $E$ is the trajectory energy obtained perhaps by a time-of -flight experiment. One finds that the effective particle state energy $\bar{E}$ is approximated by the simple expression, $$\bar{E} \approx -4 \phi \gamma^2 E \, . $$ Since $\gamma  \approx$ $ E/(m c^{2}),$ the effective energy $\bar{E}$ in the potential $\phi$ is proportional to the cube of the trajectory energy $E.$ 

Referenced to interstellar space, the potential $\phi_{E}$ at the Earth's surface  is roughly $\phi_{E} \approx$ $-1.06 \times 10^{-8},$  unitless because the potential is reduced by a factor $c^2.$ The conventional gravitational increase in kinetic energy from interstellar space to the Earth's surface is just $E_{grav}$ = $ -\phi_{E} m c^{2}$ = 10 eV. Such conventional contributions are negligible here and are ignored. For protons, the above approximation for the effective energy $\bar{E}$ is valid for $\gamma >>$ $1/\sqrt{-\phi_{E}}$ = $10^4,$ i.e. for a trajectory  energy $E$ much higher than $10^{13}$ eV, a value near the current maximum beam energy in accelerators. 

For example, a cosmic ray proton detected at the earth with an effective particle state energy of $3 \times 10^{20}$ eV, the energy of the so-called `OMG particle',\cite{OMG} would have had a particle state energy in interstellar space of about $2 \times 10^{15}$ eV, i.e. about 5 orders lower. Since the potential in interstellar space vanishes, the trajectory energy in interstellar space is also $2 \times 10^{15}$ eV.

Cosmic rays with energies exceeding $10^{20}$ eV are very rare, so for this discussion let $3 \times 10^{20}$ eV be the upper limit for the effective particle state energy at Earth, implying an upper limit of $2 \times 10^{15}$ eV for the trajectory energy which is also the upper limit for the effective particle state energy of a proton in interstellar space.

A detailed discussion of the implications of this result to cosmic ray acceleration and transport mechanisms is beyond the scope of this article. Nevertheless, some comments seem apparent. The energy in interstellar space $2 \times 10^{15}$ eV is close to the limit often cited for acceleration of cosmic rays by the interstellar shock waves of supernovae remnants. Also, the energy $2 \times 10^{15}$ eV is well under the GZK cutoff of about $6 \times 10^{19}$ eV, \cite{GZK1,GZK2} so the proton would not produce pions when interacting with Cosmic Microwave Background photons. Finally galactic magnetic fields with observed magnitudes of $2 \times 10^{-10}$ T \cite{galMag1,galMag2,galMag3} would harness a $2 \times 10^{15}$ eV proton with an orbit radius of about one parsec, so the acceleration could take place in the Galaxy. 

The model makes specific assumptions in order to explain the energies of cosmic rays detected on or near the Earth's surface. Thus the cosmic ray experiments influence the fundamental assumptions of the model and are part of the basis for the predictions of the model. The main result is the dependence of particle state energy and momentum on local gravity. The predictions of the model include the cosmic ray spectra in interstellar space and at the Sun in Fig. 2.  

\pagebreak
\section{Dual Phases} \label{Dual}

In I \cite{S1} a quantum field is constructed by modifying a well-known procedure \cite{W} giving the fields of massive free particles. For convenience the construction is briefly discussed here.

The quantum field of a particle species with nonzero mass $m$ can be constructed as a linear combination of the operators that create or annihilate the single particle states of the particle. Let $\psi_{l}(x)$ denote the quantum field and let $a_{\sigma}(\overrightarrow{p})$ and  $a_{\sigma}^{\dagger}(\overrightarrow{p})$ be the annihilation and creation operators that remove or add to a multiparticle state a single particle state of (four-)momentum $p$ and $z$-component of spin $\sigma.$ The fourth component of the momentum, the energy, can be found knowing the three spatial components $\overrightarrow{p}$ and the mass $m.$ One has 
\begin{equation} \label{psi} \psi_{l}(x) = \sum_{\sigma} \int d^{3}{\overrightarrow{p}} \, u_{l \sigma}(x,\overrightarrow{p}) a_{\sigma}(\overrightarrow{p})  + \sum_{\sigma} \int d{\overrightarrow{p}} \, v_{l \sigma}(x,\overrightarrow{p}) a_{\sigma}^{\dagger}(\overrightarrow{p})\, ,
\end{equation} 
where the $u$s and $v$s are the coefficient functions. 

Clebsch-Gordon coefficients connect quantities that transform differently with the same rotation. The coefficient functions $u$ and $v$ connect the operators $a$ and $a^{\dagger}$ with the field $\psi.$  These quantities transform differently with a Poincar\'{e} transformation, one (the operators) with unitary reps and the other (the fields) with nonunitary reps. So the $u$s and $v$s act in much the same way as Clebsch-Gordon coefficients.

A single particle state for a given momentum is a plane wave that gains a phase factor upon translation. Translations are included in Poincar\'{e} transformations of the operators  $a$ and $a^{\dagger}$ and the operators are multiplied by the same phase factor as the particle states. But the transformation of the field $\psi_{l}(x)$ does not introduce any phase factors. Thus, the coefficient functions $u$ and $v$ must contain compensating phase factors to adjust for the phase factors in the operators.

The coefficient functions $u$ and $v$ keep these compensating phase factors even when no transformations are applied. This is why a quantum field is a sum over plane waves.  

The particle states have phases and the coefficients $u$ and $v$ have phases. For free fields, the phases of the particle states and the phases of the coefficients $u$ and $v$ are the same within a sign. For the quantum fields that respond to forces in I, the phases of the particle states and the phases of the coefficient functions $u$ and $v$ differ.

To get fields that respond to forces in I, one allows the operators $a$ and $a^{\dagger}$ to transform with $(\Lambda,\epsilon)$ when the field $\psi_{l}(x)$ transforms with the Poincar\'{e} transformation $(\Lambda,\delta x).$  The operators transform by the same Lorentz transformation $\Lambda$ followed by a translation along a possibly different displacement $\epsilon.$ The basis for allowing this is the invariance of coordinate differences under translations: {\it{any translation preserves all coordinate differences.}} And coordinate differences determine the spacetime intervals whose invariance is postulated in special relativity. Since $\epsilon$ can be any displacement, it can depend on $\Lambda,$ $x,$ and $\delta x,$ i.e. $\epsilon$ is an arbitrary function $\epsilon(\Lambda,x,\delta x).$  

To show how this goes, consider a scalar annihilation field 
\begin{equation} \label{scalar1}\psi(x) = \int d^{3}{\overrightarrow{p}} \, u(x,\overrightarrow{p}) a(\overrightarrow{p}) \, .
\end{equation}
For other spins and for details, see I.
Apply the transformation $(\Lambda,\delta x)$ to the field $\psi$ and  $(\Lambda,\epsilon)$ to the operator $a,$ leaving the coefficients unchanged. One has
\begin{equation} \label{scalar2}\psi(\Lambda x +\delta x) = \int d^{3}{\overrightarrow{p}} \, u(x,\overrightarrow{p}) e^{i \Lambda p \cdot \epsilon(\Lambda,x,\delta x)} \sqrt{\frac{(\Lambda p)^{t}}{p^{t}}} a(\overrightarrow{\Lambda p})  \, ,
\end{equation}
where $p^t$ is the time component of the four-momentum.

Since the parameter $p$ in (\ref{scalar1}) is a dummy variable we can replace it with $\Lambda p.$ Then using  $d^{3}{\overrightarrow{\Lambda p}}$ = $d^{3}{\overrightarrow{p}} (\Lambda p)^{t}/p^{t},$ see Ref. \cite{3VOL}, and applying (\ref{scalar1}) with $\Lambda x + \delta x$ in place of $x,$ one has 
\begin{equation} \label{scalar3} \psi(\Lambda x + \delta x) = \int d^{3}{\overrightarrow{p}} \,\frac{(\Lambda p)^{t}}{p^{t}} u(\Lambda x +\delta x,\overrightarrow{\Lambda p}) a(\overrightarrow{\Lambda p})  \, ,
\end{equation}
implying with (\ref{scalar2}) that
\begin{equation} \label{scalar4} \sqrt{\frac{(\Lambda p)^{t}}{p^{t}}} u(\Lambda x +\delta x,\overrightarrow{\Lambda p})  = u(x,\overrightarrow{p}) e^{i \Lambda p \cdot \epsilon(\Lambda,x,\delta x)} \, .
\end{equation}
The coefficient $u$ on the left is evaluated at $\Lambda x +\delta x$ and the $u$ on the right is evaluated at $x.$ By considering a special case of (\ref{scalar4}) one can have the $u$s on both sides evaluated at $x$ = 0. 

Consider (\ref{scalar4}) with $\Lambda$ = 1 and $\delta x$ = $-x.$ This yields
 \begin{equation} \label{scalar5} u(x,\overrightarrow{ p})  = u(0,\overrightarrow{p}) e^{-i  p \cdot \epsilon(1,x, -x)} \, .
\end{equation}
This and its modification with $\Lambda x + \delta x$ in place of $x$ can be substituted in (\ref{scalar4}). One gets
\begin{equation} \label{scalar6} \sqrt{\frac{(\Lambda p)^{t}}{p^{t}}} u(0,\overrightarrow{\Lambda p})  = u(0,\overrightarrow{p}) e^{i \Lambda p \cdot [\epsilon(\Lambda,x,\delta x)-\Lambda \epsilon(1,x,-x)+\epsilon(1,\Lambda x + \delta x,-\Lambda x - \delta x)]} \, .
\end{equation}
Compare this with (\ref{scalar4}). Now both $u$s are evaluated at $x$ = 0.

The exponential in (\ref{scalar6}) cannot depend on $x$ or $\delta x,$ because nothing else in (\ref{scalar6}) does. The function  $\epsilon(\Lambda,x,\delta x)$ is thereby constrained. One can show that the displacement $\epsilon(\Lambda,x,\delta x)$ must have the following form,
\begin{equation} \label{epsilon1}
 \epsilon^{\mu}(\Lambda,x,\delta x) =  \epsilon^{\mu}(\Lambda) - \Lambda^{\mu}_{\sigma} M^{\sigma}_{\nu} x^{\nu} + {M^{\prime}}^{\mu}_{\sigma} \Lambda^{\sigma}_{\nu} x^{\nu} + {M^{\prime}}^{\mu}_{\nu}\delta x^{\nu} \, ,
\end{equation}
where $M$ = $M(x)$ is an arbitrary second rank tensor field defined over spacetime $x$ and $M^{\prime}$ = $M(\Lambda x+\delta x).$ The next section explores some consequences of (\ref{epsilon1}). 

We assume $\epsilon^{\mu}(\Lambda)$ = 0 because we are interested in the part of $\epsilon$ that is linear in $x$ or $\delta x.$ Note that one recovers $\epsilon$ = $\delta x$ when the field $M$ is the identity, $M^{\mu}_{\nu}$ = $\delta^{\mu}_{\nu}.$

The displacement $\epsilon(\Lambda,x,\delta x)$ depends on the event $x,$ so one must construct the field event-by-event, as is emphasized in I. The meaning of $M$ can be found by considering the classical trajectories of a particle described by the quantum field. This is considered next.

With expression (\ref{epsilon1}) for $\epsilon,$ one finds by (\ref{scalar5}) that
  \begin{equation} \label{u1} u(x,\overrightarrow{ p})  = u(0,\overrightarrow{p}) e^{i  p \cdot Mx} \, .
\end{equation}
Thus the coefficients $u$ are approximately `plane waves' in regions where $p$ and $M$ change slowly. By (\ref{scalar1}), $\psi$ is a sum over these `plane waves'.

Maximizing or minimizing the change in phase $\delta \Theta$ =  $p \cdot M \delta x$ determines the classical trajectory. It is shown in I that extreme phase change occurs for $M \delta x$ proportional to $p.$ Thus there is a quantity $\delta \tau $ such that 
\begin{equation} \label{tau}\frac{p^{\alpha}}{m} = M^{\alpha}_{\mu} \frac{\delta x^{\mu}}{ \delta \tau }\, .
\end{equation}
But, from the start, the momentum $p$ is in the Wigner class with constant positive mass and positive energy. The magnitude of $p$ is the mass, $p^2$ = $-m^2,$ and $p^{t} \geq m >$ 0. By (\ref{tau}) we can write $p^2$ = $-m^{2}$ as   
\begin{equation} \label{g0}\eta_{\alpha \beta} M^{\alpha}_{\mu}M^{\beta}_{\nu} \frac{\delta x^{\mu}}{\delta \tau}\frac{\delta x^{\nu}}{\delta \tau} = - 1 \, ,
\end{equation}
where $\eta_{\alpha \beta}$ is the flat spacetime metric of the spacetime upon which the field is constructed.

Define the second order tensor $g_{\mu \nu},$ 
\begin{equation} \label{g1}  g_{\mu \nu} \equiv  \eta_{\alpha \beta} M^{\alpha}_{\mu}M^{\beta}_{\nu} \, .
\end{equation}
Then, by combining the preceding three equations, we can write the equation $p^2$ = $-m^2$ as
\begin{equation} \label{g2}  g_{\mu \nu} \frac{\delta x^{\mu}}{\delta \tau}\frac{\delta x^{\nu}}{\delta \tau} = - 1 \, ,
\end{equation}
for a displacement $\delta x$ along the classical trajectory. It is shown in I that $g_{\mu \nu}$ can be taken to be the gravitational metric tensor, making $\tau$ the proper time. In what follows, we assume that $g_{\mu \nu}$ is known from general relativity and use (\ref{g1}) to restrict $M.$ Note that $M$ is not uniquely determined. At least the sign of $M$ is not determined since (\ref{g1}) is quadratic in $M.$  

It is at this point in I that one invokes a nontrivial representation of translation applied to momentum $p.$ With this assumption the trajectory of a particle is in general curved even though the spacetime $x$ is flat, because translating the momentum $p$ along a displacement changes the momentum in accordance with the nontrivial representation of translation. In general relativity, by  contrast, at least in its geometric interpretation,\cite{MTW} a flat spacetime implies there are no gravitational forces and particles move in straight lines at constant speed in the absence of other forces. Here and in I and before that in Ref. \cite{Sclassical}, particles can move along curved paths in flat spacetime. 

Note that it is the momentum $p$ that determines the trajectories by maximizing the phase $p \cdot Mx.$ Since observing the trajectory by measuring the time to travel a known distance, or some other time-of-flight method, would result in the momentum $p,$ the momentum $p$ is called the `trajectory' momentum to distinguish it from another momentum found below in Sec. \ref{s3}.

In this section, the quantum field is shown to be a sum over coefficient functions proportional to phase factors $\exp{(i  p \cdot Mx)}.$ Extreme phase change directs the particle to follow classical trajectories, as is shown in I. In the next section, the expression  (\ref{epsilon1})  for the displacement $\epsilon$ is considered, giving rise to a different phase with another momentum, an effective momentum that takes part in interactions.

\section{Cosmic Rays} \label{s3}

In this section we interpret the formula (\ref{epsilon1}) for the displacement $\epsilon$ of the operators and the particle states. The discussion leads to  the momentum of particle states which is shown to depend on gravity. Assumptions are made so that the ultrahigh energy states of some cosmic rays detected in Earth's gravity would exist in much lower energy states in interstellar space.

Let $A$ indicate the event at $x_{A}.$ Quantities in an initial reference frame are indicated by `0'. The initial coordinates and initial momentum are $x_{0}$ and $p_{0}.$ The initial coordinates of event $A$ are $x^{A}_{0}.$ After a transformation $(\Lambda,\delta x)$ the quantities are $x+\delta x,$ $p,$ and $x^{A}+\delta x.$

We can infer some properties of particle states from the way operators transform. Single particle states transform in the same way as the creation and annihilation operators that add or remove them from multiparticle states. Thus given a transformation $(\Lambda, \delta x)$ of spacetime $x_{0},$ we know that the single particle state with eigenmomentum $p_{0}$ removed by the operator $a(p_{0})$ undergoes the same transformation $(\Lambda, \epsilon^{A})$ as the operators. Here, the displacement $\epsilon^{A}$ indicates that $\epsilon$ is evaluated using (\ref{epsilon1}) at event $A.$ 

The operator $a(p_{0})$ removes the same particle state wherever the quantum field is constructed, but the transformation applied to the operator and therefore to the particle state changes from event to event.  Thus the operators are not functions of $x_{0},$ but the transformations applied to the operators are functions of $x_{0}.$

One thing that we cannot infer from the transformation of the operators is what spacetime the transformation of the particle state is applied to. Any Poincar\'{e} transformation is a symmetry operation for any spacetime coordinate system. It may be that the spacetime $y_{0}$ for a single particle state differs from the spacetime $x_{0}$ of the trajectory and quantum field. Let us assume that {\it{the transformed spacetimes differ by a Lorentz transformation}} $\lambda,$
  \begin{equation} \label{xANDy} x = \lambda y \, ,
\end{equation} 
where  $\lambda$ could be an inversion.

Thus we deduce that a single particle state created by the operator $a(p_{0})$ is a plane wave proportional to $\exp{(\pm i p_{0} \cdot y_{0})}.$ Particle states $\exp{(\pm i p_{0} \cdot y_{0})}$ and quantum fields $\psi(x_{0})$ = \linebreak$\int d^{3}p_{0} \, u(x_{0},\overrightarrow{p}_{0}) a(\overrightarrow{p}_{0})$ both have dependences on $\overrightarrow{p}_{0}$ and therefore on $p^{\mu}_{0}.$ We infer that spacetimes $x_{0}$ and $y_{0}$ must transform with Lorentz transformations just as $p^{\mu}_{0}$ transforms, so that spacetime scalar products $p_{0} \cdot x_{0}$ and $p_{0} \cdot y_{0}$ are preserved. 

By Sec. \ref{Dual}, see (\ref{scalar2}) for the operator transformation, an initial single particle state $\exp{(\pm i p_{0} \cdot y_{0})}$ transforms with the Poincar\'{e} transformation $(\Lambda,\epsilon^{A})$ when spacetime undergoes the transformation $(\Lambda,\delta x).$ The superscript $A$ indicates that the displacement $\epsilon$ depends on event $A.$ The displacement $\epsilon^{A}$ in general differs from the displacement  $\epsilon^{C}$ associated with a nearby event $C.$ 

Having different transformations act at different events motivates a closer look at how a plane wave realizes Poincar\'{e} transformations. The plane wave  assigns a phase factor $\exp{(\pm i p_{0} \cdot y_{0})}$ to an event with coordinates $y_{0}.$ The phase factor at any event, say $A,$ by itself realizes a representation of the Poincar\'{e} group. The successive transformations $(\Lambda_{1},\epsilon^{A}_{1})$ followed by $(\Lambda_{2},\epsilon^{A}_{2})$ applied to the phase factor at the event $A$ yield 
$$  e^{i p_{0} \cdot y^{A}_{0}} \rightarrow e^{i \Lambda_{1}p_{0} \cdot (\Lambda_{1}y^{A}_{0} +\epsilon^{A}_{1})}  \rightarrow e^{i \Lambda_{2}\Lambda_{1}p_{0} \cdot (\Lambda_{2}\Lambda_{1}y^{A}_{0} +\Lambda_{2}\epsilon^{A}_{1}+\epsilon^{A}_{2})} \, ,
$$
where  one recognizes the law for successive Poincar\'{e} transformations, $(\Lambda_{2},\epsilon^{A}_{2})$$(\Lambda_{1},\epsilon^{A}_{1})$ = $(\Lambda_{2}\Lambda_{1},\Lambda_{2}\epsilon^{A}_{1}+\epsilon^{A}_{2}).$

When just one transformation is applied to a plane wave, the phase factor at each event undergoes the same transformation. When there is a distinct transformation $(\Lambda,\epsilon^{A})$ at each event $A,$ a suitable generalization is to {\it{apply the distinct transformation at each event to the phase factor at that event.}} 

Therefore, the transformations $(\Lambda,\epsilon)$ are applied event-by-event to an initial plane wave $\exp{( i p_{0} \cdot y_{0})},$ yielding for the phase factor at event $A,$
  \begin{equation} \label{phase1}  e^{i p_{0} \cdot y^{A}_{0}} \rightarrow e^{i p \cdot (y^{A} +\epsilon^{A})} =  e^{i p \cdot y^{A}}e^{i p \cdot(- M + \Lambda^{-1} M \Lambda)\lambda y^{A}} e^{i p \cdot M_{0} \delta x}\, ,
\end{equation}
where the last expression follows from (\ref{epsilon1}) and (\ref{xANDy}), $p$ = $\Lambda p_{0},$ $y$ = $\Lambda y_{0},$ $M_{0}$ is the tensor in the initial reference frame and $M$ is in the transformed frame with $M_{0}$ = $\Lambda^{-1} M \Lambda $ and $M$ = $\Lambda M_{0} \Lambda^{-1}.$

Recall that $M$ is related to the gravitational metric tensor $g_{\mu \nu},$ see (\ref{g1}). For many situations with a particle moving in weak gravitational fields, the gravitational field changes little over a region of space confining the particle for a short time. One can treat $M$ as constant over fairly large regions, large on the scale of the relevent portion of the quantum field. And on such a scale, the change of the momentum $p$ is often small due to a weak gravitational force. Thus we can treat both $M$ and $p$ as constants on a scale much larger than the scale of the quantum field. 

Then the coefficient of $y^{A}$ in the phase of the exponential in (\ref{phase1}) does not depend on the event $A.$ This means that a unique, common plane wave is formed by the process. The momentum of the common plane wave is an `effective momentum' $\bar{p}$ given in
  $$  \bar{p} \cdot y = p \cdot y + p \cdot(- M + \Lambda^{-1} M \Lambda) \lambda y = p(1 - M \lambda + \Lambda^{-1} M \Lambda \lambda)\cdot y \, .
$$
The two momenta $\bar{p}$ and $p$ are equal when $\Lambda$ = 1 or $M$ = 1.

Keeping track of indices and displaying the flat spacetime metric $\eta$ yields an expression for the effective momentum. One has
  \begin{equation} \label{pbar}  \bar{p}^{\alpha} =  p^{\beta}\left[\delta^{\alpha}_{\beta} - \eta^{\alpha \rho}\eta_{\beta \tau}M^{\tau}_{\sigma}\lambda^{\sigma}_{\rho} + \eta^{\alpha \rho}\eta_{\beta \tau} (\Lambda^{-1} M \Lambda)^{\tau}_{\sigma} \lambda^{\sigma}_{\rho} \right] \, ,
\end{equation}
where summation over repeated indices is understood. Note that in general $\bar{p}$ is not the trajectory momentum $p$ = $\Lambda p_{0}$ that one would observe by measuring the time for a particle of mass $m$ to travel a known distance along the trajectory.  

Since the effective momentum $\bar{p}$ is the momentum of the plane wave obtained by transforming the particle state of momentum $p_{0},$ we assume that this plane wave forms the effective particle state in the transformed frame. Since the plane wave with the effective momentum $\bar{p}$ is the set of phase factors at each event in spacetime that describes the state of the particle, we conclude that {\it{for experiments that measure the momentum and energy of cosmic rays by the transfer of momentum and energy to other particles, it is the effective momentum and energy $\bar{p}$  that is recorded.}}

By making suitable assumptions for $M,$ $\Lambda,$  and $\lambda$ we can apply the expression for the effective momentum (\ref{pbar}) to cosmic rays so that they have less effective energy traveling in interstellar space ($M$ = 1) than they have when detected in the gravitational potential of the Earth ($M \neq$ 1). 

We only consider protons as primaries. If other particles are primaries then the results below need to be adjusted.

Consider a moving proton well-separated from other protons. The relevant portion of the proton quantum field can be confined to a tube by combining particle states with eigenmomenta spread out over a suitable range to conform with the Heisenberg Uncertainty Principle.

Spacetime outside the tube need not be translated. We assume that the external spacetime remains in a fixed frame that we call the `given frame'. We are most interested in so-called ultrahigh energy cosmic rays with energies of $10^{18}$ eV or more,  moving at nearly the speed of light relative to the Earth and Sun. So we choose the given flat spacetime reference frame to be a frame with the Earth and Sun moving with speeds negligible with respect to the speed of light. A suitably universal reference frame is provided by the distribution of Cosmic Microwave Background (CMB) radiation. The observed dipole anisotropy of the CMB implies the Solar System is moving at a speed of 370 km/s = 0.00123$c$ \cite{CMB1,CMB2} with respect to the CMB reference frame. Thus the CMB reference frame is suitable for the purposes here. We take the given frame to be the CMB reference frame.

Then the motion can be pictured as a succession of translations through displacements $\delta x_{i}$ applied to the region of spacetime inside the tube. Since the particle is confined to the tube, the resulting sequence of quantum fields is just the same as if the translations were applied to the whole of spacetime. 

Since the motion of the proton is described as a sequence of translations with displacements in various spacetime directions, the accelerations of the proton are described by translations only and without the need for rotations or boosts. Here is where the nontrivial representation of translation allows and directs the proton's accelerations. No Lorentz transformations are applied.

Now there is a problem: By (\ref{pbar}), each effective momentum $\bar{p}_{i}$ depends on an as-yet-unspecified Lorentz transformation $\Lambda_{i}.$ Thus there is a sequence of some as-yet-unspecified initial reference frames that are translated to the given frame with the as-yet-unspecified transformations $\Lambda_{i}.$ 

Lacking any reason to prefer one frame over another, we need to make an assumption before we can calculate the effective momenta $\bar{p}_{i}.$

{\it{Assumption: Each initial frame that determines each Lorentz transformation $\Lambda_{i}$ is the rest frame of the particle.}}

 Then each inital momentum ${p_{0}}_{i}$ is the momentum at rest, $k$ = $\{0,0,0,m\},$ and the Lorentz transformation $\Lambda_{i}$ is the transformation $L(p_{i})$ taking $k$ to momentum of the particle's trajectory $p_{i}$ in the CMB reference frame; see Fig. 1. We have  
 \begin{equation} \label{Lp}\Lambda_{i}k = L(p_{i})k = p_{i} \, .
\end{equation}
From here on the sequence index is dropped, e.g. $\Lambda_{i} \rightarrow $ $\Lambda$ and $p_{i} \rightarrow$ $p,$ etc.

There are many ways to transform $k$ to $p.$ For definiteness, let $L$ be
$$L^{i}_{k}(p) = \delta^{i}_{k} + (1+\gamma)^{-1} m^{-2} p^{i}p^{k}\, ,$$
 \begin{equation} \label{L} L^{i}_{4} = L^{4}_{i} = m^{-1}p^{i} \quad {\mathrm{and}} \quad L^{4}_{4} = \gamma = m^{-1} p^{t}\, ,
\end{equation}
where $m$ is the mass of the particle, i.e. the proton's mass.

We turn now to $M.$ For simplicity we assume a spherically symmetry, diagonal gravitational metric tensor $g_{\mu \nu}.$ For trajectories in a weak gravitational field with the gravitational potential $\phi$, $\phi \leq$ 0 and $\mid \phi \mid <<$ 1, one has \cite{Adler}
\begin{equation} \label{gNEWT} g_{\mu \nu} = {\mathrm{diag}}(g_{xx},g_{xx},g_{xx},g_{tt}) = {\mathrm{diag}}(1 - 2 \phi,1 - 2 \phi,1 - 2 \phi, -1 - 2 \phi) \, .
\end{equation}
By (\ref{g1}), one choice for $M$ is diagonal,  
\begin{equation} \label{Mnewt} M^{\alpha}_{\mu} = {\mathrm{diag}}(M_{x},M_{x},M_{x},M_{t}) = {\mathrm{diag}}(1 -  \phi,1 -  \phi,1 -  \phi,1 -  \phi,1 +  \phi) \, ,
\end{equation}
where terms of second order in $\phi$ are dropped. 

 The weak field gravitational potential $\phi$ at a point $Q$ in space is a sum over sources, 
\begin{equation} \label{phi} \phi = - \sum_{s} \frac{ G m_{s}}{r_{s} c^{2}} \, ,
\end{equation}
where $G$ is the universal gravitational constant, $m_{s}$ is the mass of the source  $s,$ $r_{s}$ is the spatial distance from $s$ to the point $Q,$ and $c$ is the speed of light. 

As it happens, the rotation curves for disk galaxies are flat, so the orbital speeds of stars in the disk are approximately independent of their distance from the Galactic center. There is only a small difference between the inner and outer disk kinetic energies of circularly orbiting objects (per unit mass), i.e. the potential in the disk is the same as the potential beyond the disk. Therefore, it is reasonable to assume that the gravitational potential of the Galactic disk vanishes. {\it{The gravitational potential in the Galactic disk in interstellar space far from any star is assumed to vanish.}}

Having made assumptions for $M$ and $\Lambda,$ it remains to consider $\lambda,$ which relates the spacetime for the trajectory and the spacetime for the effective particle states. If we assume the two spacetimes are the same or opposite, i.e. $ \lambda$ = $\pm 1$ in (\ref{xANDy}), then one can have the desired reduction in effective energy for the cosmic rays. But then the effective three-momentum $\overrightarrow{\bar{p}},$ the spatial part of the four-momentum $\bar{p},$ would not behave properly. 

Heuristically, one can argue as follows. The lowest order term $M$ = 1 cancels out of (\ref{pbar}). The next order term has $-\phi$ for $M_{x}$ and $+\phi$ for $M_{t},$ see (\ref{Mnewt}), making this part of $M$ proportional to the matrix diag$\{-1,-1,-1,+1\}.$ It must be that $M \lambda$ can be proportional to 1 (the delta function = diag$\{+1,+1,+1,+1\}$), since the first term in parentheses in (\ref{pbar}) is the delta function. A spatial inversion or a time inversion for $\lambda$ combined with the part of $M$ linear in $\phi$ yields a matrix $M\lambda$ proportional to 1, which can be expected to behave reasonably. One suspects that $\lambda$ could be an inversion. 
 
One finds that assuming $\lambda$ to be a time inversion produces a spatial momentum that increases when the energy increases, as one would expect. Thus, we assume that $\lambda$ {\it{is the time inversion}} 
\begin{equation} \label{lambda} \lambda^{\sigma}_{\rho} = {\mathrm{diag}}(+1,+1,+1,-1) \, .
\end{equation}
With the above assumptions for $M,$ $\Lambda,$ and $\lambda$ one can obtain a suitable expression for the effective momentum $\bar{p}.$

By (\ref{pbar}), (\ref{Lp}), (\ref{Mnewt}), and (\ref{lambda}) we have the effective momentum 
\begin{equation} \label{pbar3} \bar{p}^{\mu} = p^{\mu} (1 - 4 \gamma^2 \phi)  \, ,
\end{equation}
where only the lowest order terms in $\phi$ are kept,  $\bar{p}^{k}$ indicates the spatial part of the effective four-momentum, $k \in$ $\{1,2,3\},$ and $\bar{p}^{t}$ = $\bar{p}^{4}$ = $mc^2+\bar{E}$ is the effective total energy of the particle state, including the rest energy. Recall that $p$ is the trajectory four-momentum observable by time-of-flight measurements.

Ultra-high energy cosmic rays are observed with more energy than expected. By (\ref{pbar3}) the effective energy $\bar{p}^{t}$ is more than the energy of the trajectory since $\phi <$ 0. Because experiments determine the energy of the particle state, the measured cosmic ray energy is the quantity $\bar{E}$ = $\bar{p}^{t} - m c^{2}$ with the total energy $\bar{p}^{t}$ given in Eq. (\ref{pbar3}) with $\phi$ evaluated at the Earth.  For the accuracy needed here, the thickness of the atmosphere and the altitude of Earth orbiting satellites can be neglected.  By (\ref{phi}), we find the gravitational potential $\phi$ to be $-1.06 \times 10^{-8}$ at the Earth's surface due to the Sun and the Earth.

Experiments that measure the energy of cosmic rays have succeeded in pushing the observed spectrum to `ultra-high' energies $E > 10^{18}$ eV. Some of the data \cite{data1} - \cite{data10} is collected in the spectrum of Fig. 2 labeled `Earth'. Not all available data is included because the sketch is already busy and the discussion here involves the overall properties of the spectrum upon which all data agree.

The expected energy spectrum of the protons in interstellar space (I.S.) can be obtained from the energies of cosmic rays observed by Earth-based experiments. One applies the chain rule to obtain the predicted scaled flux in interstellar space using (\ref{pbar3}),
\begin{equation} \label{chain} \biggl(\frac{dN}{dlnE}\biggr)_{I.S.} =  \biggl(\frac{dN}{dln\bar{E}}\biggr)_{Earth} \, \frac{dln\bar{E}}{dlnE} \, ,
\end{equation}
where $(dN/dln\bar{E})_{Earth}$ is the scaled flux observed by Earth-bound experiments, $\bar{E}$ = $\bar{p}^{t} - m c^{2}$ and ${E}$ = ${p}^{t} - m c^{2}.$
The resulting energy spectrum is plotted in Fig. 2 and labelled `Interstellar Space'. 

Just as (\ref{pbar3}) and (\ref{chain}) deduce the energy spectrum in interstellar space from the observed earth-based spectrum, one can reverse the process and predict the cosmic ray spectrum for any weak potential $\phi.$ The gravitational potential at the surface of the Sun is $\phi$ = $-2.12 \times 10^{-6},$ with $\phi$ = 0 in interstellar space as before. The spectrum is plotted in Fig. 2 and labelled `Sun'. 

Note that the product of the energy $E$ and the flux $dN/dE$ is plotted in Fig. 2, $E dN/dE$ = $dN/dlnE.$ We call this the `scaled flux'. The scaled flux is closely related to the number of particles. Since the same number of particles (per unit area per unit time per steradian) are involved in each spectrum, this fact explains the alignment of the three spectra.

One sees from Fig. 2 that the spectra overlap up to about $10^{13}$ eV, where the spectra diverge and at the highest energies the spectra differ by many orders of magnitude. The predicted interstellar spectrum may be compared with the astrophysics of accelerating and transportating the protons. Such a comparison lies beyond the scope of this paper. It may be possible to detect cosmic rays striking the Sun where the highest energy cosmic rays in Fig. 2 should have energies more than a hundred times larger than those striking the Earth.

Experimental confirmation of the predicted cosmic ray spectra in interstellar space and at the Sun's surface would provide support for the assumptions and explanations presented in this article.


\begin{figure}[ht] \label{f1} 
\centering
\vspace{0cm}
\hspace{0in}\includegraphics[0,0][360,360]{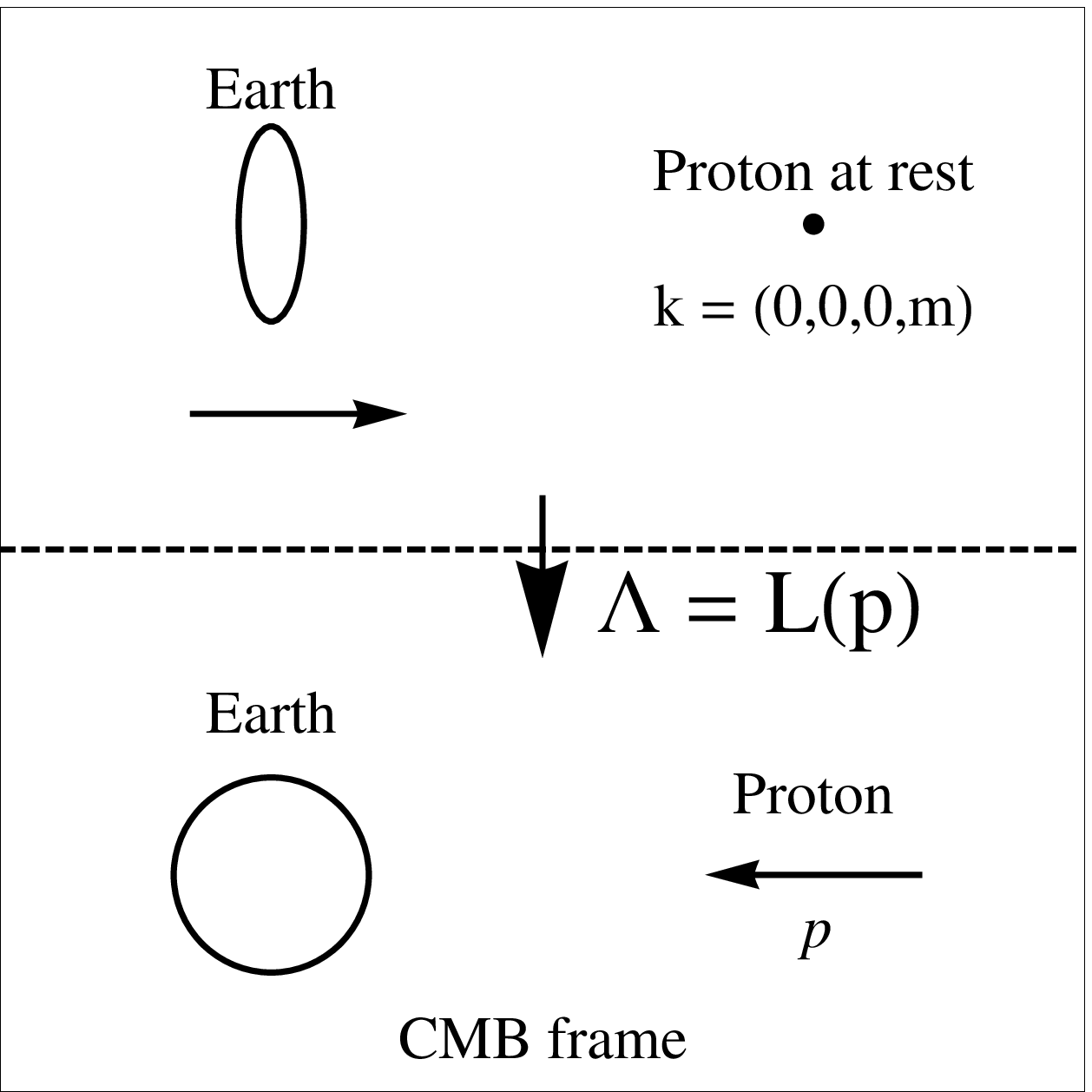}
\caption{{\it{The Lorentz Transformation $\Lambda.$}} Top: In the rest frame of the proton, the Earth is Lorentz contracted. Bottom: In the frame at rest with respect to the distribution of Cosmic Microwave Background photons, the Earth is nearly at rest and the proton has a trajectory four-momentum $p,$ as measured by  time-of-flight and mass. However the effective particle state of the proton has a different four-momentum $\bar{p}.$ The trajectory and the particle state arise from plane waves with different phases. It is $\bar{p},$ not $p,$ that is deposited into the Earth's atmosphere when the particle state changes upon interaction with the atmosphere.    }
\end{figure}

\begin{figure}[ht] \label{f2}	
\centering
\vspace{0cm}
\hspace{0in}\includegraphics[0,0][360,360]{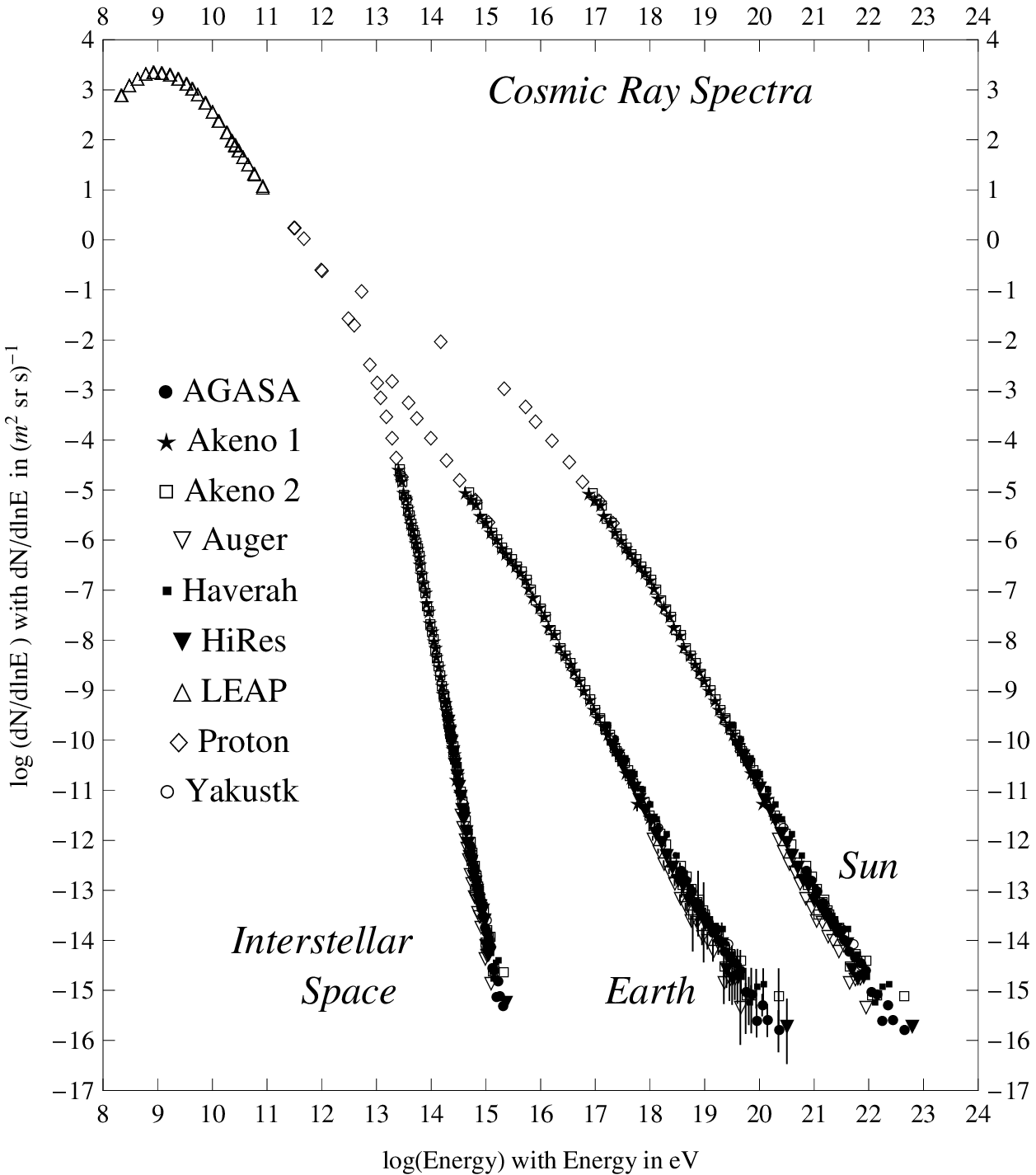}
\caption{{\it{Cosmic Ray Spectra, Dependence on Local Gravity.}} The spectrum labeled `Earth' plots the experimentally determined scaled flux incident on Earth's atmosphere.\cite{data1}-\cite{data10} The flux is scaled by the kinetic energy $E,$ i.e. $dN/dlnE$ = $E dN/dE.$ The spectrum `Interstellar Space' is the expected spectrum where the gravitational potential vanishes. For the most energetic cosmic rays, the predicted Interstellar Space spectrum is remarkably less energetic than the spectrum observed on Earth. At the Sun the magnitude of the gravitational potential is greater than at the Earth and the predicted cosmic ray energies at the Sun are higher than those detected at the Earth.} 
\end{figure}

\vspace{1.0cm}
\appendix

\pagebreak
\section{Problems} \label{Pb}

\vspace{0.3cm}
\noindent 1. A proton approaches the Earth radially with a trajectory energy of $E$ = $2 \times 10^{15}$ eV as measured in the Cosmic Microwave Background reference frame. Find the Lorentz contracted diameter of the Earth in the rest frame of the proton. 

\vspace{0.3cm}
\noindent 2. Show that if $M$ satisfies Eq. (\ref{g1}), then $A M$ also satisfies the equation with the same $g_{\mu \nu}.$ Here $A$ is an arbitrary Lorentz transformation, i.e. a sequence of rotations and boosts, but no translations.

\vspace{0.3cm}
\noindent 3. Show that the expression for $\epsilon$ in (\ref{epsilon1}) makes the phase factor in (\ref{scalar6}) independent of $x$ and $\delta x.$

\vspace{0.3cm}
\noindent 4. The spectra in Fig. 2 straighten out at high energy. (i) Fit each spectrum at high energy to a power law, 
$$ \frac{dN}{dE}  = a E^{n} \, ,  $$ where $a$ and $n$ are constants. Note that Fig. 2 plots $ EdN/dE .$ (ii) Considering an energy $E_{0}$ in the power law portion of each spectrum, use the values of $a$ and $n$ from (i) to estimate the number of cosmic rays with energy $E_{0}$ or higher and the intensity passing through one square meter in one second from each incident steradian. (iii) For Earth, compare the intensity with the Solar Constant which is about 1400 W/m$^2$. (iv) Use the results of (ii) to explain why the high energy spectra for the Earth and the Sun align as drawn in Fig. 2.

\vspace{0.3cm}
\noindent 5. Using the approximate effective momentum (\ref{pbar3}), find $dN/dE$ as a function of $dN/d\bar{E},$ $E,$ and $\phi.$

\vspace{0.3cm}
\noindent 6. The OMG particle deposited $3 \times 10^{20}$ eV of energy into the Earth's atmosphere.\cite{OMG} How much energy would the particle have deposited into the Sun's photosphere? Assume the OMG particle is a proton.

\vspace{0.3cm}
\noindent 7. Get $M$ and $\bar{E}$ for a general diagonal metric $g_{\mu \nu}.$ Assume that $M$ is diagonal and reduces to (\ref{Mnewt}) in weak fields. Then use the Schwarzchild metric in isotropic coordinates to get the energy of an incident proton at the event horizon of a collapsed object. What is the value of the energy at the event horizon for an incident proton that had an energy of $2 \times 10^{15}$ eV in interstellar space far from the collapsed object?

\vspace{0.3cm}
\noindent 8. Suppose SN A and SN B are supernovae with shock waves. Assume a proton is accelerated from shock wave A  and collides with a molecule in the atmosphere of a planet B ahead of shock wave B. A tiny fraction of the copious energy deposited into the atmosphere of planet B is used to eject a proton from planet B into interstellar space with a very low velocity.  The proton is accelerated by shock wave B and is then incident on a planet A ahead of shock wave A. The proton collides with a molecule in the atmosphere of planet A, releasing a lot of energy. Some of that energy releases protons into interstellar space to be accelerated by the shock wave A and so on. Obtain an expression for the energy gained in one such cycle.

\vspace{0.3cm}
\noindent 9. Find the ratio of a component of the particle state momentum $\bar{p}^{\mu}$ to the trajectory momentum component $p^{\mu}$ as a function of $p^{4}$ (= $p^{t}$ = $m \gamma$) and the weak gravitational potential $\phi$ using Eq. (\ref{pbar3}). Assuming the particle is a proton, graph the energy ratio $\bar{p}^{\, 4}/p^{4}$ as a function of $\log{p^{4}}$ for $p^{4}$ from 1 GeV to 10 TeV on the Earth's surface. Could experiments with currently operating proton accelerators distinguish $\bar{p}^{\, 4}$ and $p^{4}?$

\vspace{0.3cm}
\noindent 10. Find the effective mass $\bar{m}$ of the effective particle state with momentum $\bar{p}$ in (\ref{pbar3}) as a function of both the gravitational potential $\phi$ and the trajectory gamma $\gamma$ = $p^{t}/mc^{2}.$ Plot $\bar{m}$ vs $\gamma$ in interstellar space, at the Earth's surface, and on the surface of the Sun. 

\vspace{0.3cm}
\noindent 11. Derive an expression like (\ref{pbar3}) except assume that the spacetime for particle states is the negative of the trajectory spacetime, i.e. $\lambda$ = $-1.$ What happens to the spectra in Fig. 2? What happens to the effective three-momentum $\overrightarrow{\bar{p}}?$ What happens to the effective mass $\bar{m}$ in Problem 10?


\section{Figures} \label{figs}

\end{document}